\documentclass[11pt]{article}
\usepackage{graphicx}
\usepackage{amsmath,amsfonts,amssymb,bm}
\newcommand{\Z}{\mathbb Z}
\newcommand{\St}{\mathsf S}

\newcommand{\Eq}[1]{Eq.~(\ref{#1})}
\newcommand{\Fig}[1]{Fig.~\ref{fig:#1}}
\newcommand{\Sec}[1]{Sec.~\ref{sec:#1}}

\title{On generalization of reversible second-order 
cellular automata}
\date{\today}
\author{Alexander Yu.\ Vlasov}
\begin{document}
\sloppy
\maketitle
\begin{abstract}
 A cellular automaton with $n$ states may be used for construction
 of reversible second-order cellular automaton with $n^2$ states. 
 Reversible cellular automata with hidden parameters
 discussed in this paper are generalization of such construction 
 and may have number of states $N=n m$ with arbitrary $m$. Further
 modification produces reversible cellular automata with 
 reduced number of states $N' < N = n m$. 
\end{abstract}

\section{Introduction}

{\em Second-order} cellular automaton (CA) may be used for construction of 
reversible CA \cite{TM90,WCA}. The approach let us construct reversible
CA with set of states $\mathsf D = \St \times \St \simeq \Z_{n^2}$ from arbitrary
CA with states from some $\St \simeq \Z_n$. In the paper is suggested extension
of such idea with $\mathsf D = \St \times \mathsf H \simeq \Z_{nm}$
with further reduction to some $\mathsf R \subset \mathsf D$.
Here instead of second term $\St$ corresponding to previous
state is used arbitrary set $\mathsf H$ considered as 
hidden parameter.

Some basic structures are introduced in \Sec{pre}, yet a 
preliminary acquaintance with properties of CA is supposed. 
The construction of reversible second-order CA is revisited in
\Sec{scnd}. Generalizations $\mathsf D = \St \times \mathsf H$ and
$\mathsf R \subset \mathsf D$ are introduced
in \Sec{gen} and \Sec{red} respectively. Some examples are 
presented for clarification in \Sec{eg}.

\section{Preliminaries}
\label{sec:pre}

A {\em space of cells} $X$ is defined here in a general way as
{\em a finite or infinite discrete set}. Simple examples of such spaces are
$\Z^d$ and some subset of such lattices. Element $x \in X$ 
is called here {\em location} (of cell).

Set of states of each cell is a finite discrete
set $\St$ with $n$ elements. A map 
$I_{\St} : {\St} \rightarrow  \Z_n$ 
enumerates the states by indexes $0 \ldots n-1$.

Complete description of each cell is a pair 
$(s,x)$ also denoted here as $(s)_x$ with $x \in X$,
$s \in \St$.
{\em Configuration} of whole CA is a map $\St^X : X \rightarrow \St$.

For a space of cells such as $X = \Z^d$ the neighborhood of given
cell $\vec{x} \in \Z^d$ with $j$ {\em adjacent cells} 
is often described via shifts 
$(\vec{x}+\vec{\nu_1},\dots,\vec{x}+\vec{\nu_j})$ \cite{WCA,KT13}, 
but for more general $X$ it should be defined 
as some function $\nu : X \rightarrow X^j$.

{\em Local update rule} is a function 
$f : \St^{j+1} \rightarrow \St$, 
where $(s,s_1,\ldots,s_j) \in \St^{j+1}$ is the vector 
of states for a neighborhood together with the cell itself.
For a cell $(s)_x$ the notation $f[s]_x$ or $f[s]$  
may be used for $f(s,s_1,\ldots,s_j)$.

\medskip

Consideration of cellular automata on Penrose's tilings 
\cite{PenTile} demonstrates that such a definition of neighborhood 
and local update rules may be still not enough to describe some models.
The problem is not only varying number $j_x$ of adjacent cells, but also
a few possible geometries and orientations of the neighborhood.

To include such models, let us use formal notation $[\St]$ for
{\em a domain of definition} of local update rule.
Element of such set is denoted here as $[s] \in [\St]$. 
In simplest case discussed above $[\St] = \St^{j+1}$,
$[s] = (s,s_1,\ldots,s_j)$. 
For variable size of neighborhood it may be defined 
$[\St] = \bigcup_k \St^{j_k+1}$, 
$[s] = \bigl(k,(s,s_1,\ldots,s_{j_k})\bigr)$. 

For any location $x$ and each configuration $\St^X$
{\em localization} is a map 
\begin{equation}
\Xi : (\St^X,x) \rightarrow [\St].
\label{LocConf}
\end{equation}
The value of such a map for given $x$ is designated
further as $[s]_x$ and for local update function
may be used notation $f[s]$ or $f[s]_x$ already 
introduced earlier.

In simpler and more regular cases localization $\Xi$ of
configuration $\St^X$ may be defined as composition of two maps. 
First map $\nu$ gives neighboring cells for given $x$. Localization 
$\St^{j+1}=[\St]$ is a restriction of configuration
$\St^X$ to given neighborhood.

\section{Second-order cellular automata}
\label{sec:scnd}

Let us consider arbitrary CA with set of states 
$\St = \Z_n$ and local update rule $f$. 
Second-order CA \cite{TM90} produced
from such a rule has the same set of locations $X$,
but the state is a pair from set 
$\mathsf D = \St \times \St$ 
with $n^2$ elements and update rule is
\begin{equation}
f_R : (s,s') \mapsto (f[s]\ominus s', s),
\label{rf}
\end{equation}
where `$\ominus$' is subtraction modulo $n$. 

The update rule \Eq{rf} may be decomposed into
two reversible steps, {\em an update:}
\begin{equation}
f_u : (s,s') \mapsto (s,f[s]\ominus s' )
\label{fu}
\end{equation}
and {\em the swap:}
\begin{equation}
w : (s,s'') \mapsto (s'',s).
\label{swp}
\end{equation}
The rule $f_R = w\,f_u$ is reversible 
$f_R^{-1} = f_u^{-1} w^{-1}$,
where $w^{-1} = w$, $f_u^{-1} = f_u$.

\medskip

The global reversibility of local operation $w$ \Eq{swp} is rather 
obvious, because it acts on each cell separately. Operation
$f_u$ \Eq{fu} uses values of neighboring cells, but 
it acts only on second parameter of the state. Due to 
construction of the $f_u$ \Eq{fu} neighboring cells are
not affected by the second parameter (before application of $w$)
 and such a property produces global reversibility 
of $f_u$ in spite of the fact that definition \Eq{fu} is local.

\medskip

Generalization for less regular spaces $X$ with different configurations
and number of neighboring cells is rather straightforward with
{\em localization} $\Xi$ of configuration ${\mathsf D}^X$ \Eq{LocConf}.
In fact, it is enough to consider {\em localization} $\Xi$ of ${\St}^X$ to 
construct $f_u$ \Eq{fu} and swap $w$ only acts on cell itself.

\section{Generalization with hidden parameters}
\label{sec:gen}

The term {\em second-order}  describes possibility
to use information about previous state \cite{TM90}.
In local update rule such as \Eq{rf} the information is
accessible only to cell itself. 
It is possible to consider yet another interpretation 
of such update rules. The set of states is direct product
of two components, but only one is accessible for neighboring
cells.

Such alternative approach may be used for generalizations,
if to consider the second component not as previous state,
but as an arbitrary hidden parameter of a cell inaccessible 
for neighbors. 
In such a case size of both components may be different and 
instead of update \Eq{fu} and 
exchange \Eq{swp} could be used more general functions. 

\smallskip

Let us consider CA with states from set of pairs 
$\mathsf D = \St \times \mathsf H$, $\St = \Z_n$, 
$\mathsf H = \Z_m$ and update rule represented as
composition of two steps
\begin{equation} 
f_{D} = \varpi f_\upsilon.
\label{fD}
\end{equation} 
Here $\varpi$ is a fixed reversible transformation of $\mathsf D$ 
and 
\begin{equation}
f_\upsilon : (s,h) \mapsto \bigl(s,F[s](h)\bigr),
\label{ups}
\end{equation}
where $F[s](h) = F(s,s_1,\ldots,s_j)(h)$ must be reversible 
function on set $\mathsf H$ for any vector of states
$(s,s_1,\ldots,s_j)$. 

In fact, not only second-order CA is particular
case of such extension for $\mathsf H = \St$, but also arbitrary
$n$-order CA may be considered in such a way with $\mathsf H = \St^{n-1}$
storing $n-1$ previous states and $\mathsf D = \St \times \mathsf H = \St^n$.

In most general case,
reversible functions on $\mathsf H$ may be represented via 
transpositions $\pi_m$ of $m$ elements of $H$ and so 
instead of $f: \St^{j+1} \rightarrow \St$ ({\em cf} usual CA 
recollected in \Sec{pre}) should be used map $F$
\begin{equation}
F : \St^{j+1} \rightarrow \pi_m.
\label{Fgen}
\end{equation}
Only for set with two states both $\St = \Z_2$ and $\pi_2$ 
have two elements and such distinction is not essential.

Unlike the usual second-order CA discussed in \Sec{scnd}
such an extended version is not necessary
directly derived from some CA. On the other hand, it may be useful sometimes to start
with a second-order CA and to consider local update rule with bigger number 
of output states \Eq{Fgen}. Simple example is discussed further in \Sec{ex3}
for CA with three states.

\medskip

Construction for spaces $X$ with variable number
of neighboring cells is also straightforward due to
{\em localization} $\Xi$ of configuration ${\mathsf D}^X$ \Eq{LocConf}.
It is again enough to consider {\em localization} of ${\St}^X$ to construct
$F$ \Eq{Fgen} and $\varpi$ affects only state of cell itself.

\section{Reduced number of states} 
\label{sec:red}

Set with $nm$ states used above may be redundant.
Reduction of previous construction on some subset 
$\mathsf R \subset \St \times \mathsf H$ is discussed below.

Let us consider some discrete set $\mathsf R$ with two projections
$p_1 : \mathsf R \rightarrow \St$ 
and $p_2 : \mathsf R \rightarrow \mathsf H$. So, each
state $r \in \mathsf R$ corresponds to pair 
$r \simeq \bigl(p_1(r),p_2(r)\bigr)$.

\begin{figure}[htb]
\begin{center}
\includegraphics[scale=0.5]{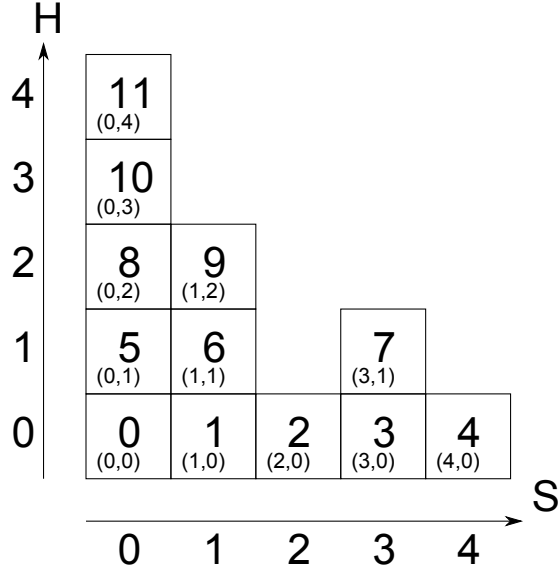}
\end{center}
\caption{Scheme with reduced set of states} 
\label{fig:reduc}
\end{figure}

As an example, \Fig{reduc} with 12 states may be 
represented as set of pairs: {\small 
$\{(0,0),(1,0),(2,0),(3,0),(4,0),\,
(0,1),(1,1),(1,3),\,(0,2),(1,2),\,(0,3),\,(0,4)\}.$}
\smallskip

Let us denote $m(s)$ number of states $r$ with $p_1(r)=s$, {\em e.g.}, 
$m(0) = 5$, $m(1) = 3$, $m(2)=m(4)=1$, $m(3)=2$ for \Fig{reduc}.

Due to representation of states from $\mathsf R$ via set of pairs
it is possible again to use two-steps rule such as \Eq{fD} with
$\varpi$ is transposition on $\mathsf R$ and
$f_\upsilon$ is an analogue of \Eq{ups}  
\begin{equation}
f_\upsilon : \bigl(p_1(r),p_2(r)\bigr) \mapsto 
 \bigl(p_1(r),F[p_1(r)](p_2(r))\bigr),
\label{redups}
\end{equation}
where $F[p_1(r)]=F(p_1(r),p_1(r_1),\ldots,p_1(r_j))$
is reversible function (transposition) 
on set with $m(p_1(r))$ states. 
For example on \Fig{reduc} function $F[0]$ may exchange only 
five states $\{0,5,8,10,11\}$, $F[1]$ --- $\{1,6,9\}$, {\em etc}.
The $F$ depends only on $p_1$ projections of states in
the neighborhood and rearranges states of the cell 
without change of $p_1$.

There is a formal difficulty in description of $F$ in \Eq{redups}. 
A map such as \Eq{Fgen} may not be used directly, because value 
$m$ in $\pi_m$ for a state $r \in \mathsf R$ depends on 
$p_1(r) = s \in \St$.

\begin{figure}[htb]
\begin{center}
\includegraphics[scale=0.5]{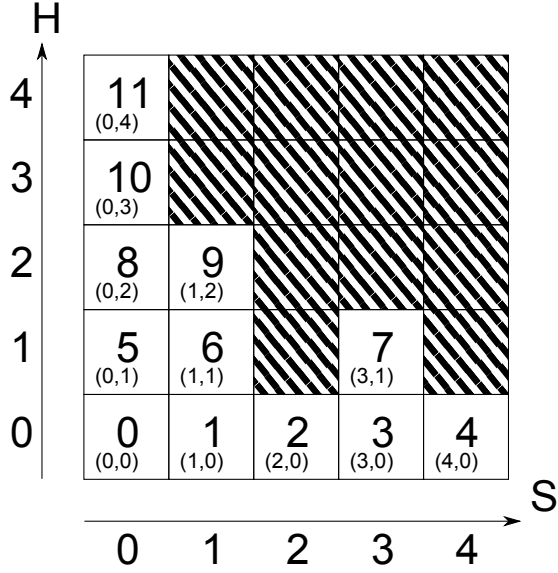}
\end{center}
\caption{Reduced set of states $\mathsf R$ included into 
$\St \times \mathsf H$} 
\label{fig:reduc_in}
\end{figure}

A simple way to resolve 
such a problem --- is to consider bigger formal set of states 
$\St \times \mathsf H \supset \mathsf R$, \Fig{reduc_in}, but to use 
functions $F$ and $\varpi$ changing only states from $\mathsf R$. In such
a case it is possible to use 
$F : \St^{j+1} \rightarrow \pi_{m_{\max}}$, where $m_{\max} \geq m(s)$
for any $s$.

Another method is to consider {\em a family} of functions 
\begin{equation}
 F_s : \St^j \rightarrow \pi_{m(s)}.
\label{Fred}  
\end{equation}
Here $\St^j$ is used instead of $\St^{j+1}$, because
state of cell itself formally is not used as an argument
of a function $F_s$ and is treated in a special way as an index 
inside of the family.
 
It is not necessary to introduce formal set $\St \times \mathsf H$
with auxiliary states in such a case.
Such approach may also require an insignificant change of notation
such as $F[s]=F_s(s_1,\ldots,s_j)$ instead of $F[s]=F(s,s_1,\ldots,s_j)$.

\medskip

The construction of CA with variable number of neighboring cell 
may be again carried out with {\em localization} $\Xi$ of configuration 
${\mathsf R}^X$ or $\St^X$ \Eq{LocConf}. 
It works both with formal consideration of $\mathsf R$ as a
reduction of auxiliary space $\St \times \mathsf H$ and with 
definition of $\mathsf R$ via family of functions $F_s$.

\section{Examples}
\label{sec:eg}

\subsection{CA with \bm{$2\times 2 = 4$} states}
\label{sec:ex2}

\begin{figure}[htb]
\begin{center}
\includegraphics[scale=0.5]{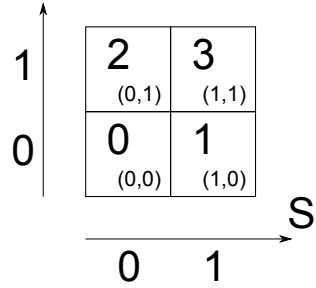}
\end{center}
\caption{Second order CA from CA with two states} 
\label{fig:ca2x2}
\end{figure}

Let us consider two-states cellular automaton with
transition function $f$. For two states 
there is no difference between subtraction and addition modulo
two, the same binary operation is also known as {\sf XOR}
(e{\sf X}clusive {\sf OR}) and \Eq{rf} may be written in 
more familiar way:
\begin{equation}
f_R : (s,s') \mapsto (f[s] \mathop{\sf XOR} s', s).
\label{rf2}
\end{equation}
The \Eq{rf2} is transition function for reversible cellular
automaton with four states \Fig{ca2x2}. 

The local transition function of initial CA may have only two values
and they realize two possible reversible function on set with two
elements \Eq{rf2}
\begin{equation}
  \begin{array}{|c|c|c|}\hline
  f & s' & f \mathop{\sf XOR} s' \\ \hline \hline
  0 & 0 & 0 \\
  0 & 1 & 1 \\ \hline
  1 & 0 & 1 \\
  1 & 1 & 0 \\ \hline 
  \end{array}
\label{r2tab} 
\end{equation}

With lack of other generalizations of function $f_\upsilon$ \Eq{ups} for 
two states only $\varpi$ may be changed.

\subsection{CA with \bm{$3\times 3 = 9$} states}
\label{sec:ex3}

\begin{figure}[htb]
\begin{center}
\includegraphics[scale=0.5]{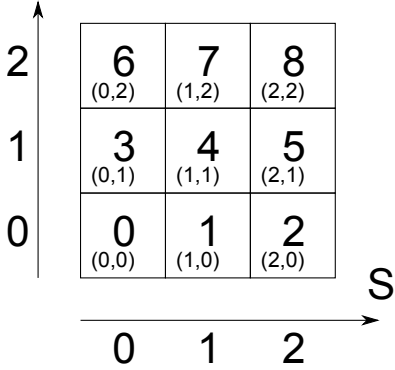}
\end{center}
\caption{Second order CA from CA with three states} 
\label{fig:ca3x3}
\end{figure}

Let us consider for comparison initial CA with three states. 
\begin{equation}
f_R : (s,s') \mapsto (f[s] + 3 - s' \bmod 3, s).
\label{rf3}
\end{equation}
The \Eq{rf3} is transition function for reversible cellular
automaton with nine states \Fig{ca3x3}. 

Only three between six reversible transformations ($3! = 6$) are
represented in \Eq{rf3} and so it is possible to consider extensions
with six different $f_\upsilon$ \Eq{ups}. For three states all six
possible functions may be written quite naturally, {\em e.g.}
\begin{subequations}
 \label{ups3}
\begin{align}
 f_0(h) & = h \\
 f_1(h) & = h + 1 \bmod 3\\
 f_2(h) & = h + 2 \bmod 3\\
 f_3(h) & = 3 - h \bmod 3\\
 f_4(h) & = 4 - h \bmod 3\\
 f_5(h) & = 2 - h.
\end{align}
\end{subequations}

It is clear, that only $f_3$, $f_4$ and $f_5$ are used in \Eq{rf3}.
Even without change of `swap' function $\varpi$ there are possible extensions for
such CA, due to possibility to associate six different actions \Eq{ups3} 
with local update rule instead of three.
The inverse for $f_3$, $f_4$, $f_5$ and $f_0$ --- is the function itself and
$f^{-1}_1 = f_2$.

\subsection{Example of $\bm{\varpi}$ for CA with \bm{$3\times 2 = 6$} states}
\label{sec:ex23}

\begin{figure}[htb]
\begin{center}
\includegraphics[scale=0.5]{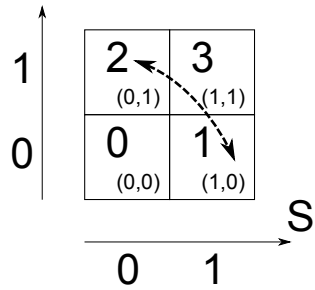}
\end{center} 
\caption{Swap operation $\varpi$ for $\mathsf H = \St$} 
\label{fig:ca2x2sw}
\end{figure}

In examples above the operation $\varpi$ in \Eq{fD} may simply exchange two components
of the state  because $\mathsf H = \St$.
Such a swap for CA with $2 \times 2$ states is shown on \Fig{ca2x2sw}. 
 
\begin{figure}[htb]
\begin{center}
\includegraphics[scale=0.5]{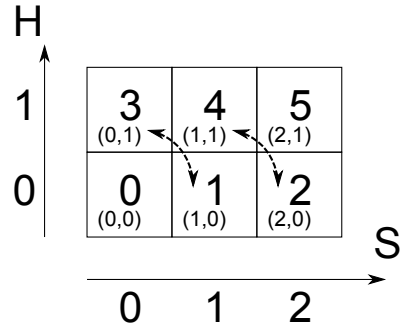}
\end{center} 
\caption{Operation $\varpi$ for $3\times 2$ CA} 
\label{fig:ca3x2sw}
\end{figure}

Let us consider $3\times 2$ CA with six states, where
$\St = (0,1,2)$, $\mathsf H = (0,1)$. An example
of operation $\varpi$ for such CA is shown on \Fig{ca3x2sw}.

\subsection{CA with ``blank'' state}
\label{sec:ex5}

An example of application of constructions discussed
in \Sec{red} is considered below. Let us start with some
standard or extended second-order CA based on two-states CA such 
as \Sec{ex2}.

It is useful sometimes to add special ``blank'' cells. 
A cellular space of CA may be restricted to some domain
of arbitrary shape if to set all complementary cells to such
``blank'' value. Such a state may not be changed, but
other states must be updated by the update function of 
initial CA without ``blank'' state.

\begin{figure}[htb]
\begin{center}
\includegraphics[scale=0.5]{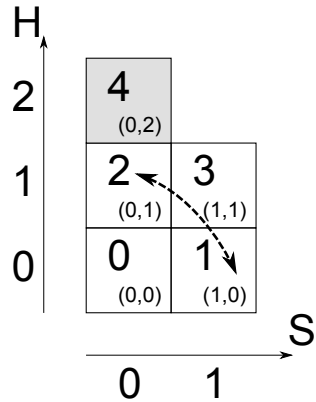}
\end{center}
\caption{CA with ``blank'' state derived from second-order $2\times 2$ CA} 
\label{fig:ca32sw}
\end{figure}

On the \Fig{ca32sw} is shown extension of $2\times 2$ CA with
``blank'' state (4). The function $\varpi$ is denoted by
arrows and exchanges only two states $(1\leftrightarrow 2)$, the 
function $f_\upsilon$ should be derived from initial CA and may 
change only first four states.

It should be mentioned, that term ``reduced number of states'' here 
should be treated with some care --- given example could be formally described 
as reduction of CA with $6 = 2\times 3$ states, but it would not reflect
properly idea of ``blank'' cells used for construction of the CA.

\end{document}